\documentclass[a4paper,11pt]{article}
\pdfoutput=1 

\usepackage{jheppub} 

\usepackage[T1]{fontenc} 

\title{\boldmath The Effect of Rényi Entropy on Hawking Radiation}

\author{Yang Liu}

\affiliation[a]{School of Physics and Astronomy, University of Nottingham, Nottingham NG7 2RD, UK}
\affiliation[b]{Nottingham Centre of Gravity, University of Nottingham, Nottingham NG7 2RD, UK}

\emailAdd{yang.liu@nottingham.ac.uk}

\abstract{It is  widely believed that Hawking radiation originates from excitations near the horizons of black holes [1,2,3]. However, Giddings [2] proposed that the Hawking radiation spectrum that characterizes evaporating semi-classical black holes originates from a quantum “atmosphere”, which extends beyond the horizon of a black hole. Although several research projects have been conducted in this field, they have not yet taken into account the effect of Rényi entropy. In the present article, we will therefore consider the effect of Rényi entropy on Hawking radiation power.  We assume that if the effect of Rényi entropy is very small, we suggest that the Hawking radiation should originate from the quantum “atmosphere” which extends beyond the black hole's horizon for finite dimensions. That is, that Giddings' suggestion is the more likely of the above possibilities. However, for infinite dimensions, both suggestions are equally credible. We briefly consider the very large effect of Rényi entropy on Hawking radiation power as well. We find that if the effect of Rényi entropy is very large and $\omega/T_{BH}$ is very small, then the power spectral density $S_R$ is proportional to the power spectral density $S_{BH}$.}

\begin{document} 
\maketitle
\flushbottom

\section{Introduction}
\label{sec:intro}

Based on Hawking's original paper, the evaporation process of black holes is characterized by a non-unitary evolution of quantum fields in curved spacetimes [1]. More specifically, matter fields in a pure quantum state will collapse to form a black hole that eventually evaporates into a mixed thermal state [1]. It is widely believed that the semi-classical Hawking radiation  spectra should be modified to ensure quantum unitarity since unitary temporal evolution is one of the fundamental principles  of quantum mechanics [2,3]. \\
It is generally believed that Hawking radiation spectra originate from quantum excitations near the horizon region, namely, $\Delta r = r - r_H \ll r_H$ [2,3]. Therefore, it was widely expected that  modified Hawking radiation spectra should also be characterized by a relatively short length scale $\Delta r \ll r_H $, where $r_H$ is the radius of the black hole´s horizon [3].\\
However, Giddings suggested that the radiation spectra of a  black hole originates from an effective quantum “atmosphere” which extends well beyond the horizon of the black hole [2].  By comparing the numerical results for the Hawking radiation power $P_{BH}$ of an evaporating $(3 +1)$-dimensional Schwarzschild black hole of horizon radius $r_H$ with the familiar Stefan–Boltzmann radiation power $P_{BB} = \sigma A T^4$ of a $(3+1)$-dimensional flat space perfect blackbody emitter of radius $r_A$ [4,5], Giddings concluded that the source of Hawking radiation was a quantum region which was located outside the black hole and that the effective radius $r_A$ was described by the following relation
\begin{equation}\label{eq:1.1}
\Delta r = r_A - r_H \sim r_H. 
\end{equation}
According to ref.[2], eq.$(1.1)$ is consistent with the existence of an effective emitting “atmosphere” which extends well beyond the horizon of a black hole, i.e., “atmosphere” is outside and far from the horizon of a black hole.\\ 
In considering the correctness of Giddings' conclusion, one interesting question that should be addressed is if the radiation spectra of a black hole in any $(D + 1)$-dimension originates from an effective quantum “atmosphere” which extends well beyond the horizon of the black hole? In order to answer this question, in ref.[6], the author studied the Hawking radiation spectra of Schwarzschild black holes in a $(D+1)$-dimension. Following Giddings's method, the author defined the effective radii $r_A(D)$ of the black-hole quantum atmospheres by equating the Hawking radiation powers of the $(D +1)$-dimensional black holes to the corresponding Stefan–Boltzmann radiation powers of flat space perfect blackbody emitters [6]. The dimensionless radii $r_A/r_H$ parameters, which characterize the effective black-hole quantum atmospheres,  are functions which monotonically decrease as  the number $(D + 1)$ increases [6]. In particular, in the large $D \gg 1$ regime, ref.[6] suggests that radiating $(D +1)$-dimensional Schwarzschild black holes are characterized by the relation $(r_A - r_H)/r_H\ll 1$.\\
Following Giddings' argument, other studies on black-hole quantum atmospheres have been conducted [7,8,9]. For example, in ref.[7], Yan-Gang Miao and Zhen-Ming Xu investigated the Hawking radiation cascade from a five-dimensional charged black hole with a scalar field coupled to higher-order Euler densities in a conformally invariant manner. Myungseok Eune and Wontae Kim [8] showed that the outer temperature vanishes at the horizon and has a peak at a scale whose radial extent is set by the horizon radius which  then decreases to the Hawking temperature at infinity. Ramit Dey, Stefano Liberati and Daniele Pranzetti [9] have shown  that the Hawking quanta originate from a quantum atmosphere around the black hole with energy density and fluxes of particles peaking at about $4MG$, which is contrary to the popular belief that these quanta originate from ultra high energy excitations very close to the horizon. At present, the theory of a “black-hole quantum atmosphere” is a frontier area of research which is still far from reaching any final widely accepted conclusion. Results on this topic  which are not directly related to the present article are not listed here for reasons of brevity.\\
One important element that appears to have received little attention in previous articles is the following. In 1902 Gibbs pointed out that, in a system where the partition function diverges, the standard Boltzmann–Gibbs theory cannot be applied, and large-scale gravitational systems are known to fall within this class [10]. Then Tsallis generalized standard statistical mechanics (which arises from the hypothesis of weak probabilistic correlations and their connection to ergodicity) to nonextensive one, which can be applied in all cases, and still possessing standard Boltzmann–Gibbs theory as a limit [10]. Therefore, the usual Boltzmann–Gibbs additive entropy must be generalized to the nonextensive, i.e., non-additive entropy (the entropy of the whole system is not necessarily the sum of the entropies of its sub-systems), which is named Tsallis entropy [10]. Non-extensive statistical mechanics does not only include one example, i.e., Tsallis entropy. Rényi entropy is another main example of such a non-extensive entropy. In this paper, we will therefore principally discuss the effect of Rényi entropy on Hawking radiation power, in particular on the effective radius of a Schwarzschild black hole quantum atmosphere. In section 2, we introduce the basic concepts of Hawking radiation spectra of Schwarzschild black holes and Rényi entropy. In section 3, we derive the effective radius of a black-hole quantum atmosphere taking Rényi entropy into consideration. In section 4, numerical results of the effective radius of the black-hole quantum atmosphere are  obtained. In section 5, we briefly discuss when the effect of Rényi entropy is very large, how it impacts on Hawking radiation power. In section 6, we discuss the results we have obtained.\\

\section{Basics}
\subsection{The Hawking radiation spectra of (D+1)-dimensional Schwarzschild black holes}
For $(D+1)$-dimensional Schwarzschild black holes, the semi-classical Hawking radiation power for one bosonic degree of freedom is described as [4,5,11,12]
\begin{equation}\label{eq:2.1}
P_{BH} = \frac{1}{2^{D-1} \pi^{D/2} \Gamma(D/2)} \sum_j \int_{0}^{\infty} \Gamma \frac{\omega^D}{\exp(\omega/T_{BH}) -1} d\omega, 
\end{equation}
Here $j$ is the angular harmonic index of the emitted field modes and $\omega$ is the emitted frequency of field. The Bekenstein-Hawking temperature of the black hole is given by 
\begin{equation}\label{eq:2.2}
T_{BH} = \frac{(D-2)}{4\pi r_{BH}}.
\end{equation}
We have set $G=c=k_B=\hbar=1$. $r_{BH}$ is the horizon radius of the black hole [6,13]. The greybody factors of the black-hole-field system composition are denoted as $\Gamma=\Gamma (\omega;j,D)$, which are dimensionless coefficients which quantify the interaction of the emitted fields with curved black-hole spacetime [4,5]. 

\subsection{An area law prescription for Rényi entropy}
One of the most remarkable discoveries in fundamental physics is that the entropy of a black hole has a value equal to a quarter of the horizon area in Planck units [3,14]:
\begin{equation}\label{eq:2.3}
S= \frac{Area(Horizon)}{4G_N}.
\end{equation}
Here $G_N$ is the Newton's constant. Ryu and Takayanagi [14,15] generalized the relationship eq.$(2.3)$ in the context of gauge/gravity duality. They proposed that the von Neumann entropy is determined by the area of a codimensional-2 minimal surface in the dual spacetime [14,15]:
\begin{equation}\label{eq:2.4}
S= \frac{Area(Minimal \, Surface)}{4G_N}.
\end{equation}
The above discussions of area laws eq.$(2.3)$ and $(2.4)$ are limited to the von Neumann entropy. However, according to Gibbs' argument [10], the Boltzmann-Gibbs (BG) theory cannot be applied in systems with divergence in the partition function, such as in a  gravitational system [10]. Therefore, thermodynamic entropy of such non-standard systems cannot be described by an extensive entropy but must instead be generalized to a non-extensive entropy [10]. In ref.[14], the author has generalized the area laws to the case of Rényi entropy, in which a label index $q$ is used and the entropy is defined in terms of the density matrix $\rho$ of the entangling region as [14]:
\begin{equation}\label{eq:2.5}
S_q= \frac{1}{1-q} \ln Tr \rho^q.
\end{equation}
When $q \rightarrow 1$, the von Neumann entropy $S =- Tr(\rho \ln \rho)$ is recovered. Rényi entropy contains richer physical information about the entanglement structure of a quantum state. The properties of Rényi entropy have been extensively studied using by numerical methods [16], in spin chains [17], in tensor networks [18], in free field theories [19], in two-dimensional CFTs [20] or higher [21], and in the context of gauge/gravity duality [22]. These results have also been generalized to charged [23] and supersymmetric cases [24].\\
The generalized area-law prescription for holographic Rényi entropy can be derived by applying the replica trick in the context of gauge/gravity duality [14,25]. Rényi entropy $(2.5)$ of integer $q > 1$ is determined by the partition function of the QFT on a branched cover $M_q$, defined by taking $q$ copies of the original Euclidean spacetime $M_1$ on which the QFT lives with a cut along the entangling region and gluing them along the cuts in a cyclic order. This can be written as [14,25]:
\begin{equation}\label{eq:2.6}
S_q= \frac{1}{1-q} (\ln Z[M_q] - q \ln Z[M_1]),
\end{equation}
where $Z[M_1]$ and $Z[M_q]$ are the partition functions on the branched original and cover spacetime, respectively. For holographic QFTs, one can calculate $Z[M_q]$ by finding the dominant bulk solution $B_q$ whose asymptotic boundary is $M_q$ [14,25]. In the large $N$ limit where the bulk physics is classical, we have [14]:
\begin{equation}\label{eq:2.7}
Z[M_q] = e^{-I_{bulk}[B_q]},
\end{equation}
where $I_{bulk}[B_q]$ denotes the on-shell action of the bulk solution $B_q$. If we assume the $\mathbb{Z}_q$ replica symmetry, then one can obtain that [14]:
\begin{equation}\label{eq:2.8}
S_q = \frac{q}{q-1} (I_{bulk}[\hat{B}_q] - I_{bulk}[\hat{B}_1]), 
\end{equation}
where $\hat{B}_q = B_q / \mathbb{Z}_q$ and the bulk action $I_{bulk}$ is the Einstein action 
\begin{equation}\label{eq:2.9}
I_{bulk} = - \frac{1}{16 \pi G_N} \int d^{d+1} X \sqrt{G} R + I_{matter}.
\end{equation}
Here $X^{\mu}$, $G_{\mu\nu}$ and $R$ are the coordinates, metric and Ricci scalar in the bulk [14]. As a result, one can obtain a generalized area law:
\begin{equation}\label{eq:2.10}
q^2 \partial_q (\frac{q-1}{q} S_q) = \frac{Area(Cosmic \, Brane_q)}{4G_N},
\end{equation}
where the cosmic brane is analogous to the Ryu-Takayanagi minimal surface [14]. The above arguments can be generalized to non-integer $q$ [14]. More details can be found in ref.[14].\\ 

\subsection{Basic formulae for Rényi entropy}
A long-standing problem of non-extensive thermodynamics is deriving a formulation which is compatible with the zeroth law of thermodynamics [26]. Based on the concept of composability, Abe showed [26,27] that the most general non-additive entropy composition rule which is compatible with a homogeneous equilibrium has the form
\begin{equation}\label{eq:2.11}
H_\lambda (S_{12})= H_\lambda (S_{1})+H_\lambda (S_{2}) +\lambda H_\lambda (S_{1}) H_\lambda (S_{2}),
\end{equation}
where $H_\lambda$ is a differentiable function of $S$ and $\lambda$ is a real constant parameter. $S_1$, $S_2$ and $S_{12}$ are the entropies of the subsystems and the total system, respectively. \\
Biró and Ván [28] showed that, the most general entropy function  compatible with the zeroth law for homogeneous systems has the following form:
\begin{equation}\label{eq:2.12}
L(S) = \frac{1}{\lambda} \ln[1+ \lambda H_\lambda (S)],
\end{equation} 
which is additive for composition, namely,
\begin{equation}\label{eq:2.13}
L(S_{12}) = L(S_1) + L(S_2).
\end{equation}
The corresponding temperature is then
\begin{equation}\label{eq:2.14}
\frac{1}{T} = \frac{\partial L(E)}{\partial E},
\end{equation}
where we have assumed the principle of additivity holds for the energy composition [26]. \\
For classical black holes, the Bekenstein-Hawking formula satisfies the eq.$(2.3)$, i.e.,
\begin{equation}\label{eq:2.15}
S_{12} = S_1 + S_2 + 2 \sqrt{S_1} \sqrt{S_2},
\end{equation}
where $H_\lambda (S) = \sqrt{S}$ and $\lambda \rightarrow 0$ [26,27]. \\
The Rényi entropy can be defined as  
\begin{equation}\label{eq:2.16}
S_R = \frac{1}{1-q} \ln \sum_i p^q_i,
\end{equation}
which is equivalent to the choices of $H_\lambda (S) = S$ and $\lambda=1-q$ in eq.$(2.4)$ [26], if the original entropy functions satisfy the following condition
\begin{equation}\label{eq:2.17}
S_{12} = S_1 + S_2 +\lambda  S_1  S_2,
\end{equation} 
where $q$ is a real parameter called the non-extensivity parameter. It can be shown that the formal logarithm of Tsallis entropy 
\begin{equation}\label{eq:2.18}
S_T = \frac{1}{1-q} \sum_i (p^q_i-p_i),
\end{equation}
provides the Rényi entropy, namely,
\begin{equation}\label{eq:2.19}
S_R = L(S_T)= \frac{1}{1-q} \ln[1+ (1-q) S_T].
\end{equation}
When $q$ approaches to 1 $(\lambda \rightarrow 0)$, both Rényi and Tsallis entropy give the Shannon entropy $S=-\sum_i p_i \ln p_i$ [26]. 

\section{The effective radius of a black-hole quantum atmosphere taking Rényi entropy into consideration}
In this section, in order to determine the location of effective radius of a Schwarzchild black hole, we will derive the formulae for black hole thermodynamics considering the perturbations of Rényi entropy.

\subsection{The Rényi temperature of a (D+1)-dimensional Schwarzschild black hole}
Considering eq.$(2.2)$, $T_{BH} = M'(r_{BH})/S'(r_{BH}) = \frac{(D-2)}{4\pi r_{BH}}$, we have
\begin{equation}\label{eq:3.1}
S(r_{BH})=\frac{4\pi}{(D-2)} \int r_{BH} M'(r_{BH}) dr_{BH}. 
\end{equation}
Based on ref.[6], the horizon radius of a $(D+1)$-dimensional Schwarzschild black hole of mass $M$ is given by
\begin{equation}\label{eq:3.2}
r_{BH} = [16 \pi M / (D-1) \hat{A}_{D-1}]^{1/(D-2)}, 
\end{equation}
where $\hat{A}_{D-1} = 2 \pi^{D/2} / \Gamma (D/2) $ is the generalized area of a unit $(D-1)$-sphere. Then we have
\begin{equation}\label{eq:3.3}
M'(r_{BH}) = \frac{(D-1) \pi^{D/2}}{8\pi \Gamma (D/2)} (D-2) r^{D-3}_{BH}.
\end{equation} 
Then the thermodynamic entropy \footnote{The reason why we emphasize “thermodynamic” is that this formula is derived by thermodynamics absolutely without using the definitions of entropy in statistical mechanics.} of $(D+1)$-dimensional Schwarzschild black hole is
\begin{equation}\label{eq:3.4}
S(r_{BH}) = \frac{4\pi}{(D-2)} \int r_{BH} M'(r_{BH}) dr_{BH} = \frac{\pi^{D/2}}{2 \Gamma (D/2)} r^{D-1}_{BH}. 
\end{equation}
We have pointed out that the entropy of large-scale gravitational system cannot be described by Gibbs-Boltzmann statistical mechanics and should be replaced by Tsallis entropy [10]. Then if we regard the entropy of a black hole as being Tsallis entropy, namely, $S(r_{BH}) = S_T$ [26,29], then the Rényi entropy of a $(D+1)$-dimensional Schwarzschild black hole is then
\begin{equation}\label{eq:3.5}
S_R = \frac{1}{\lambda} \ln(1 + \lambda S(r_{BH}))  = \frac{1}{\lambda} \ln(1 + \lambda \frac{\pi^{D/2}}{2 \Gamma (D/2)} r^{D-1}_{BH}), 
\end{equation}
where $\lambda = 1-q$. We can therefore conclude that irrespective of the value of $\lambda$, the value of $S_R$ will be positive. The derivative of Rényi entropy can then be obtained as
\begin{equation}\label{eq:3.6}
S'_R = \frac{1}{1 +  \lambda \frac{\pi^{D/2}}{2 \Gamma (D/2)} r^{D-1}_{BH} }  \frac{\pi^{D/2}}{2 \Gamma (D/2)} (D-1) r^{D-2}_{BH}.
\end{equation}
So that the Rényi temperature \footnote{Here Rényi temperature refers to the black-hole temperature considering the effect of Rényi entropy.} of a $(D+1)$-dimensional Schwarzschild black hole is
\begin{equation}\label{eq:3.7}
T_R = M'(r_{BH})/S'_R= \frac{(D-2)}{4\pi r_{BH}} + \lambda \frac{(D-2) \pi^{D/2}}{8 \pi \Gamma (D/2)} r^{D-2}_{BH}.   
\end{equation}
When $D=3$ and $\lambda \rightarrow 0$, the Rényi temperature becomes
\begin{equation}\label{eq:3.8}
T = \frac{1}{4\pi r_{BH}},   
\end{equation}
which is the well-known result for a $(3+1)$-dimensional Schwarzschild black hole [6].\\ 
Eq.$(3.7)$ can be rewritten as $T_R = T_{BH} (1+x) $, where $x = \lambda \frac{\pi^{D/2} r^{D-1}_{BH}}{2\Gamma (D/2)}$. In section 3, we consider only the case of $|x| \ll 1$ and a modification to order $x$.

\subsection{The effective radius of a black-hole quantum atmosphere taking Rényi entropy into account}
On the one hand, for $(D+1)$-dimensional Schwarzschild black holes, the semi-classical Hawking radiation power for one bosonic degree of freedom should be modified as
\begin{equation}\label{eq:3.9}
P_R = \frac{1}{2^{D-1} \pi^{D/2} \Gamma(D/2)} \sum_j \int_{0}^{\infty} \Gamma \frac{\omega^D}{\exp(\omega/T_R) -1} d\omega, 
\end{equation}
where $T_R = T_{BH} (1+x) $ and $x = \lambda \frac{\pi^{D/2} r^{D-1}_{BH}}{2\Gamma (D/2)}$. Since $|x| \ll 1$, then $\frac{1}{\exp(\omega/T_R)-1}$ is approximately equal to 
\begin{equation}\label{eq:3.10}
\frac{1}{\exp(\omega/T_R)-1} \approx \frac{1}{\exp(\omega/T_{BH})-1} + \frac{\frac{x \omega}{T_{BH}} \exp(\omega/T_{BH}) }{[\exp(\omega/T_{BH})-1]^2}.
\end{equation}
So that eq.$(3.9)$ can be written as
\begin{equation}\label{eq:3.11}
\begin{aligned}
P_R =& \frac{1}{2^{D-1} \pi^{D/2} \Gamma(D/2)} \sum_j \int_{0}^{\infty} \Gamma \frac{\omega^D}{\exp(\omega/T_{BH}) -1} d\omega\\
&+ \frac{1}{2^{D-1} \pi^{D/2} \Gamma(D/2)} \sum_j \int_{0}^{\infty} \Gamma \omega^D \frac{ \frac{x \omega}{T_{BH}} \exp(\omega/T_{BH}) }{[\exp(\omega/T_{BH})-1]^2} d\omega. 
\end{aligned}
\end{equation}
The first term in eq.$(3.11)$ is the original Hawking radiation power. The second term of eq.$(3.11)$ can be written as
\begin{equation}\label{eq:3.12}
-x \beta_{BH} \frac{\partial}{\partial \beta_{BH}} P_{BH},
\end{equation}
where $\beta_{BH} = 1/T_{BH}$. Then eq.$(3.11)$ is equal to 
\begin{equation}\label{eq:3.13}
P_R=P_{BH} -x \beta_{BH} \frac{\partial}{\partial \beta_{BH}} P_{BH}.
\end{equation}
On the other hand, the scalar radiation power of a spherically-symmetric blackbody (BB) at temperature $T$ and having radius $R$ in $(D+1)$ spacetime dimensions is given by the generalized Stefan-Boltzmann radiation law [6]
\begin{equation}\label{eq:3.14}
 P_{BB} = \sigma A_{D-1} (R) T^{D+1},
\end{equation} 
where $\sigma$ is the Stefan-Boltzmann constant in $(D+1)$-dimensions:
\begin{equation}\label{eq:3.15}
 \sigma = \frac{D \Gamma (D/2) \zeta(D+1)}{2 \pi^{D/2+1} },
\end{equation}
and $A_{D-1} (R)$ is the surface area of an emitting body in $(D+1)$-dimensions:
\begin{equation}\label{eq:3.16}
A_{D-1} (R) = \frac{2 \pi^{D/2}}{\Gamma (D/2)} R^{D-1}.
\end{equation} 
In this article, we suggest that the Stefan-Boltzmann law $(3.14)$ still holds since it is a non-gravitational system which can be described by the standard statistical mechanics. However, it should be modified as,
\begin{equation}\label{eq:3.17}
P_{BB} = \sigma A_{D-1} (r) T^{D+1}_R,
\end{equation}
where $T_R$ and $r$ are the Rényi temperature and radius in $(D+1)$-dimensions, respectively. The reason can be explained as the following. If we repeat the derivation of Stefan-Boltzmann's law [30], we can know that the only thermodynamic quantity is temperature $T$, while other physical quantities are mechanical quantities. Now that we have considered the effect of Rényi entropy and Rényi temperature, the only change in Stefan-Boltzmann's law is to change the ordinary temperature $T$ into Rényi temperature $T_R$, and the other parts remain unchanged. \\
Giddings pointed out that the effective radius of a black-hole atmosphere can be determined by equating the radiation power $P_{BB}$ of a flat space perfect blackbody emitter with the Hawking radiation power $P_{BH}$ [2,6]. Therefore, following ref.[2,6], we can define the effective radius $r_A$ of the black-hole quantum atmosphere by the following relation:
\begin{equation}\label{eq:3.18}
P_{BH} (r_{BH}, T_{BH}) = P_{BB} (r_A, T_{BH}).
\end{equation}  
In this article, the eq.$(3.18)$ is modified to,
\begin{equation}\label{eq:3.19}
P_R (r_{BH}, T_R(T_{BH})) = P_{BB} (r_A, T_R(T_{BH})),
\end{equation}
where $T_R(T_{BH})$ means that $T_R$ is a function of $T_{BH}$.\\
Since $T_R=T_{BH}(1+x)$ and $|x| \ll 1$, therefore
\begin{equation}\label{eq:3.20}
T^{D+1}_R = T^{D+1}_{BH} (1+x)^{D+1} \approx T^{D+1}_{BH} [1+(D+1)x] =  T^{D+1}_{BH} + (D+1)x T^{D+1}_{BH},
\end{equation}
where we have used the binomial theorem in the second step.\\
Assuming now, that eq.$(3.18)$ still holds and considering eq.$(3.13)$, $(3.18)$, $(3.19)$ as well as $(3.20)$, we then have,
\begin{equation}\label{eq:3.21}
 \sigma A_{D-1} (r_A) (D+1) x T^{D+1}_{BH} = -x \beta_{BH} \frac{\partial}{\partial \beta_{BH}} P_{BH} (r_{BH}, T_{BH}).
\end{equation}
From eq.$(3.21)$, we have,
\begin{equation}\label{eq:3.22}
T^{D+1}_{BH} = C_{D+1} P_{BH} (r_{BH},T_{BH}),
\end{equation}
where $C_{D+1}$ is a constant in $(D+1)$-dimensions. Considering eq.$(3.14)$ and $(3.18)$, we can therefore derive that,
\begin{equation}\label{eq:3.23}
C_{D+1} = 1/\sigma_D A_{D-1} (r_A),
\end{equation}
which means that our assumption is self-consistent.\\
Combining eq.$(3.13)$, $(3.17)$, $(3.19)$ and $(3.22)$, we can obtain the effective radius $r_{AR}$ of a black-hole quantum atmosphere as being,
\begin{equation}\label{eq:3.24}
r_{AR} = \{\frac{\pi}{D \zeta (D+1)} (\frac{4\pi}{D-2})^{D+1} \bar{P}_{BH}\}^{\frac{1}{D-1}} \times r_{BH} \times \{\frac{\frac{D+1-x}{D+1}}{1+(D+1)x} \}^{\frac{1}{D-1}},
\end{equation}
where we have defined the scaled Hawking radiation power of a black hole in $(D+1)$-dimensions [6] as,
\begin{equation}\label{eq:3.25}
\bar{P}_{BH} = P_{BH} \times r^2_{BH}.
\end{equation}
If we define the effective radius $r_A$ of the black-hole quantum atmosphere without taking into account the effect of Rényi entropy [6] as 
\begin{equation}\label{eq:3.26}
r_A = \{\frac{\pi}{D \zeta (D+1)} (\frac{4\pi}{D-2})^{D+1} \bar{P}_{BH}\}^{\frac{1}{D-1}} \times r_{BH},  
\end{equation}
and the entropy difference as,
\begin{equation}\label{eq:3.27}
F = \{\frac{\frac{D+1-x}{D+1}}{1+(D+1)x} \}^{\frac{1}{D-1}}.
\end{equation}
When $x = 0$, then $F =1$, and the effective radius $r_{AR}$ of the black-hole quantum atmosphere, taking into account the effect of Rényi entropy, can be written as 
\begin{equation}\label{eq:3.28}
r_{AR} = r_A F.
\end{equation}
There are two reasons why we only expand $\frac{1}{\exp(\omega/T_R)-1}$ to order $x$. Firstly, if we expand the term to order $x$ then the effective radius $r_{AR}$ can be written as the product of the standard effective radius $r_A$ and the difference $F$, which makes for an easier comparison of the results with that given by standard statistical mechanics. Secondly, standard statistical mechanics can describe the universe accurately, therefore, the effect of Rényi entropy can be neglected completely under normal circumstances. In fact, as we have pointed out in section 2.3, when $x$ is zero, the standard results are recovered. \\

\section{Numerical results for the effective radius of a black-hole quantum atmosphere}
Following ref.[6], we can define the functional dependence $\bar{r}_{AR} = \bar{r}_{AR} (D)$ of the dimensionless ratio
\begin{equation}\label{eq:4.1}
\bar{r}_{AR} \equiv (r_{AR} - r_{BH}) / r_{BH},
\end{equation}
which characterizes the effective quantum atmospheres of the radiating black holes in $(D+1)$ dimensions. Considering eq.$(3.28)$, we can rewrite eq.$(4.1)$ as,
\begin{equation}\label{eq:4.2}
\bar{r}_{AR} = \tilde{r}_A F -1,
\end{equation}
where $\tilde{r}_A = r_A/r_{BH}$. Based on Hawking's and Giddings' arguments [2,6], there are two possible values of $\bar{r}_{AR}$ in $(D+1)$ dimensions, namely, $\bar{r}_{AR} \sim O(0)$ or $\bar{r}_{AR} \sim O(1)$. We will calculate the numerical results for the two possibilities respectively.\\
For $\bar{r}_{AR} \sim O(0)$, if we set $\bar{r}_{AR} = 0$, we have, 
\begin{equation}\label{eq:4.3}
x_D = \frac{(1-\tilde{r}^{1-D}_A)(D+1)}{(D+1)^2 \tilde{r}^{1-D}_A +1}.
\end{equation}
For $\bar{r}_{AR} \sim O(1)$, if we set $\bar{r}_{AR} = 1$, we have, 
\begin{equation}\label{eq:4.4}
x_D = \frac{(1-2^{D-1}\tilde{r}^{1-D}_A)(D+1)}{2^{D-1}(D+1)^2 \tilde{r}^{1-D}_A +1},
\end{equation}
where the subscript $D$ denotes the dimensionality of space.\\

\subsection{The (3 +1)-dimensional case}
The Hawking radiation power of scalar quanta from a $(3+1)$-dimensional Schwarzschild black hole is given by [6,31]
\begin{equation}\label{eq:4.5}
P_{BH}(D=3)= 2.976 \times 10^{-4} \frac{1}{r^2_{BH}}. 
\end{equation}
Then we have $\tilde{r}_A = 2.679$ for the effective radius of the black-hole quantum atmosphere. \\
For $\bar{r}_{AR} \sim O(0)$, from eq.$(4.3)$, we have $x_3= 1.0661$, which is $O(1)$ and inconsistent with the assumption in section 3, namely, $|x| \ll 1$. We cannot adjust the value of $\bar{r}_{AR}$ to make $|x| \ll 1$. Therefore, for the $(3+1)$-dimensional case, Hawking's suggestion is less feasible.\\
For $\bar{r}_{AR} \sim O(1)$, from eq.$(4.4)$, we have $x_3= 0.1786$. In fact, we can adjust the value of $\bar{r}_{AR}$ so that $x$ can be very closed to zero. Therefore, for the $(3+1)$-dimensional case, Giddings' suggestion is more feasible.\\

\subsection{The (D +1)-dimensional cases}
In the previous subsection, we have seen that the effective quantum atmosphere for a $(3 +1)$-dimensional Schwarzschild black hole is described by the relation
\begin{equation}\label{eq:4.6}
\bar{r}_{AR} \equiv (r_{AR} - r_{BH}) / r_{BH}.
\end{equation}
Based on the numerical results in $(3+1)$ dimensions, we can deduce, that Hawking's suggestion is less feasible than that of Giddings'. \\
We shall now calculate the values of the dimensionless radii $\bar{r}_{AR}$, which describe the effective quantum atmospheres of the radiating Schwarzschild black holes, and the value of $x$, which quantifies the effect of Rényi entropy in $(D+1)$-dimensions. In ref.[6,32], the Hawking radiation powers of Schwarzschild black holes in $(D+1)$-dimensions have been obtained numerically. We list the results of the dimensionless radii $r_A(D)$ in Table 1 [6].\\
\begin{table}
	\caption{The dimensionless radii $r_A(D) \equiv (r_A - r_{BH})/r_{BH} $ and $\tilde{r}_A = r_A/ r_{BH}$ which characterize the effective quantum atmospheres of the radiating black holes in $(D+1)$ dimensions.}
	\begin{tabular}{|c c c c c c c c|} 
		\hline
		D+1 & 5 & 6 & 7 & 8 & 9 & 10 & 11 \\
		\hline
		$(r_A- r_{BH})/r_{BH}$ & 0.982 & 0.727 & 0.590 & 0.502 & 0.439 & 0.391 & 0.355 \\
		\hline
        $\tilde{r}_A = r_A/ r_{BH}$ & 1.982 & 1.727 & 1.590 & 1.502 & 1.439 & 1.391 & 1.355 \\
        \hline
     \end{tabular}    
\end{table}

\subsubsection{The Hawking's cases for intermediate D-values}
For $\bar{r}_{AR} \sim O(0)$, from eq.$(4.3)$ and Table 1, we have $x_4 = 1.0349$. Similarly, we can obtain the values for other dimensions. The results have been listed in Table 2. Based on Table 2, we find that if $\bar{r}_{AR} \sim O(0)$, then the values of $x_D$ cannot be $O(0)$, therefore we can conclude that for any dimensions, Hawking's suggestion does not appear to be feasible.\\
\begin{table}
	\caption{The values of $x_D$ for the radiating black holes in $(D+1)$ dimensions using Hawking's suggestion.}
	\begin{tabular}{|c c c c c c c c|} 
		\hline
		D+1 & 5 & 6 & 7 & 8 & 9 & 10 & 11 \\
		\hline
		$x_D$ & 1.0349 & 1.0552 & 1.0841 & 1.1110 & 1.1302 & 1.1416 & 1.1612 \\
		\hline
	\end{tabular}    
\end{table}\\

\subsubsection{The Giddings' cases for intermediate D-values}
For $\bar{r}_{AR} \sim O(1)$, from eq.$(4.4)$ and Table 1, we have $x_4 = -5.152 \times 10^{-3}$. Similarly, we can obtain the values for other dimensions. The results have been listed in Table 3. Based on Table 3, we find that if $\bar{r}_{AR} \sim O(1)$, then the values of $x_D$ can be very closed to zero and we can therefore conclude that for any dimensions, Giddings' suggestion appears to be feasible.\\
\begin{table}
	\caption{The values of $x_D$ for the radiating black holes in $(D+1)$ dimensions using Giddings' suggestion.}
	\begin{tabular}{|c c c c c c c c|} 
		\hline
		D+1 & 5 & 6 & 7 & 8 & 9 & 10 & 11 \\
		\hline
		$x_D$ & $-5.152 \times 10^{-3} $ & -0.0729 &-0.0969 &-0.1023 &-0.0990 &-0.0945 &-0.0882 \\
		\hline 
	\end{tabular}   
\end{table}\\
Combining this with the conclusions in section 4.1 and 4.2.1, we can deduce that if we consider the effect of Rényi entropy, for any dimensions, Hawking's suggestions appear to be much less feasible that those of Gidding. Therefore, Hawking radiation most likely  originates from a  quantum atmosphere which extends beyond the horizons of black holes. We can also infer that the above conclusion is a general result and true independently of the value of $\lambda$ in non-extensive statistical mechanics.

\subsubsection{The large D regime}
In this section, we will consider the case for in the large $D$ regime to see whether the conclusions which we have obtained in section 4.1, 4.2.1 and 4.2.2 are still established. The Hawking radiation spectrum for a Schwarzschild black holes in $(D+1)$ dimensions is characterized by $\omega^D / (\exp (\omega / T_R) -1)$ (see eq.$(3.9)$). This frequency dependent function has a peak at [6]
\begin{equation}\label{eq:4.7}
\frac{\omega_{peak}}{T_R} = D + W(-D e^{-D}),
\end{equation}
where $W(x)$ is the Lambert function and $T_R$ is the Rényi temperature of the black hole as given in eq.$(3.7)$. Inserting eq.$(3.7)$ into eq.$(4.7)$, we then have
\begin{equation}\label{eq:4.8}
\frac{\lambda_{peak}}{r_{BH}} (1 + x) \approx \frac{8 \pi^2}{D^2} [1 + O(D^{-1})] \ll 1,
\end{equation}
where $\lambda_{peak}$ is the peak of wave length. Since $|x| \ll 1$, the modification term does not affect the above conclusion, namely, that in the large $D$ regime the characteristic wavelength is much shorter than the horizon radius $r_{BH}$ of a radiating Schwarzschild black hole, which is consistent with the result obtained without taking Rényi entropy into consideration.\\
Ref.[6] and [32] have pointed out that the inequality $(4.8)$ implies that the corresponding Hawking emission spectra of these higher-dimensional black holes are described extremely well by a short wavelength approximation. In particular, the effective radiating $r_A$ of a black hole in the large $D$ regime is determined by the high-energy (short wavelengths) absorptive radius of the black hole [6,12,32,33,34,35,36,37,38], namely,
\begin{equation}\label{eq:4.9}
\tilde{r}_A \equiv r_A / r_{BH} = (D/2)^{\frac{1}{D-2}} \sqrt{\frac{D}{D-2}}.
\end{equation} 
In the large $D \gg 1$ regime, we find that ,
\begin{equation}\label{eq:4.10}
\tilde{r}_A \equiv r_A / r_{BH} = 1 + O(\frac{\ln D}{D}) \approx 1.
\end{equation} 
For $\bar{r}_{AR} \sim O(0)$, $F \approx 1$. Considering eq.$(4.3)$, we obtain that $(D+1)x \approx 0$, which is the necessary condition for Hawking's arguments.\\
For $\bar{r}_{AR} \sim O(1)$, $F \approx 2$. Considering eq.$(4.4)$, we obtain that $\frac{1}{1+(D+1)x} \approx 2^{D-1}$. For large $D$, then $(D+1)x \approx -1$, which is the necessary condition for Giddings' arguments. We also find that $\lambda < 0$.\\
Since $D \gg 1$, we know that $|x|$ is extremely small for the two cases, so that the effect of Rényi entropy can be neglected completely. In addition, these two suggestions are both feasible.  

\section{Very large effect of Rényi entropy on Hawking radiation power}
In this section,  we  will briefly mention the properties of Hawking radiation power in the situation where the effect of Rényi entropy is extremely large. Although this is a situation  which does not exist in the world we have explored, it is nevertheless of interest.\\
On the one hand, since $T_R = T_{BH} (1+x)$, if $|x| \gg 1$, then $T_R = x T_{BH}$. We then have
\begin{equation}\label{eq:5.1}
\exp (\omega/T_R) -1 \approx \frac{\omega}{x T_{BH}}.
\end{equation}
Then the power spectral density of $P_R$ is
\begin{equation}\label{eq:5.2}
S_R = dP_R/d\omega = \frac{x T_{BH}}{2^{D-1} \pi^{D/2} \Gamma(D/2)} \sum_j \Gamma(\omega;j,D) \omega^{D-1},
\end{equation}
where $\Gamma(\omega;j,D)$ are the greybody factors of the black-hole field system composition. \\
On the other hand, if $\omega/T_{BH} \ll 1$, then 
\begin{equation}\label{eq:5.3}
\exp(\omega/T_{BH}) -1 \approx \frac{\omega}{T_{BH}}.
\end{equation} 
The power spectral density of $P_{BH}$ is
\begin{equation}\label{eq:5.4}
S_{BH} = dP_{BH}/d\omega = \frac{T_{BH}}{2^{D-1} \pi^{D/2} \Gamma(D/2)} \sum_j \Gamma(\omega;j,D) \omega^{D-1}.
\end{equation}
Therefore, if $x \gg 1$ and $\frac{\omega}{T_{BH}} \ll 1$, then we obtain
\begin{equation}\label{eq:5.5}
S_R = x S_{BH}.
\end{equation}
The meaning of this relationship and the effect of Rényi entropy on Hawking radiation should be studied in further research.

\section{Summary and discussion}
It is widely believed that Hawking radiation originates from the excitations near the horizons of black holes. Therefore, most researchers assume that the modified Hawking radiation spectra should also be characterized by a relatively short length scale $\Delta r \equiv r_A - r_{BH} \ll r_{BH} $ [2,3]. One of the main arguments for the above assumption is based on Hawking's calculation, involving tracing back the modes all the way from future infinity to the past null infinity, through collapsing matter, so that one has a vacuum state near the horizon for a free-falling observer [9].\\
However, Giddings [2] has proposed that the Hawking radiation spectrum that characterizes evaporating semi-classical black holes originates from an quantum “atmosphere”, which extends beyond the horizon of the black hole. Giddings has provided evidence that, for a Schwarzschild black hole in $(3 +1)$ dimensions, the source of the Hawking radiation is a quantum region outside the black-hole horizon whose effective radius $r_A$ is characterized by the relation $\Delta r \sim r_H $ [2]. Following Giddings' argument, other studies on black-hole
quantum atmospheres have been conducted [7,8,9]. Ref.[6] raised an interesting question, namely, whether Giddings' argument is of a general nature and is true for all radiating black holes irrespective of their dimensions? The results of ref.[6] showed that at least in some cases, the effective radii of the black-hole quantum atmospheres are characterized by the relation $r_A \ll 1$. This conclusion implies that the Hawking radiation originates from quantum excitations very near the horizon of a black hole [6].\\
However,  the following important consideration appears not to have  been taken into account in these previous articles. Based on Gibbs' argument [10], the Boltzmann-Gibbs (BG) theory cannot be applied to systems with divergence in the partition function, such as in a gravitational system. Therefore, the thermodynamic entropy of such non-standard systems cannot be described merely by an extensive entropy but must instead be generalized to a non-extensive entropy of which Rényi entropy is one of the main examples.\\ 
This article therefore addressed the question in ref.[6] concerning the effect of Rényi entropy. The general structure of the present article is similar to that of ref.[6]. In section 3, we derived the entropy and temperature of black holes taking Rényi entropy into consideration. The expressions are universally valid. Considering the fact that even if there indeed exists the effect of Rényi entropy, it must be very small since for now Gibbs-Boltzmann statistical mechanics can still describe physical phenomenon accurately. Therefore, we further obtained the expression for the effective radius of a quantum atmosphere including a modification term of order $x$ ($x$ is very small, i.e., $\lambda = 1-q$ is very small and $q$ approaches to 1).\\
In section 4, we obtained numerical results for the effective radius of a black-hole quantum atmosphere in $(3+1)$ dimensions and in $(D+1)$ dimensions. Based on our calculated  results listed in Tables 2 and 3, we suggested that it is much more feasible that Hawking radiation originates from a quantum “atmosphere” which extends beyond the horizons of black holes having finite dimensions as suggested by Giddings. Furthermore, if we want to see whether this conclusion is still established, we found that for black holes having infinite dimensions, however, two suggestions are both feasible. In section 5, we briefly considered also the very large effect of Rényi entropy on Hawking radiation power. If the effect of Rényi entropy is very large and $\omega/T_{BH}$ is very small, we found that the power spectral density $S_R$ is proportional to $S_{BH}$, that is, $S_R =x S_{BH}$, where $x$ is defined as in section 3.1 and includes the effect of Rényi entropy.


\end{document}